%% file: approx-tp.tex
\documentclass[preprint]{elsarticle}
\usepackage[mathlines]{lineno}
\usepackage{bbm}
\usepackage{enumerate}
\usepackage{booktabs}
\usepackage{tabu}
\usepackage[ruled,lined,boxed,linesnumbered]{algorithm2e}
\usepackage{amsfonts}
\usepackage{xcolor}
\usepackage{xspace}
\usepackage{float}
\usepackage{mathtools}
\usepackage{tabularx}
\usepackage{relsize}
\usepackage{amsmath}
\usepackage{amssymb}
\usepackage{amsthm}
\usepackage{bm}
\usepackage{cleveref}
\usepackage[main=english,czech]{babel}

\biboptions{sort&compress}

\theoremstyle{plain}
\newtheorem{theorem}{Theorem}

\newtheorem{lemma}{Lemma}
\newtheorem{observation}{Observation}
\newtheorem{corollary}{Corollary}

\newtheorem{Reduction Rule}{Reduction Rule}
\newtheorem{Claim}{Claim}
\newcommand{\lstpair}{{local $s$-$t$ pair}\xspace}

\newcommand{\stpath}{{$s$-$t$ path}\xspace}
\newcommand{\stpaths}{{$s$-$t$ paths}\xspace}

\newcommand{\tprob}{{\sc Tracking Paths}\xspace}
\newcommand{\fvsp}{{\sc Feedback Vertex Set}\xspace}
\newcommand{\FVS}{{\sc FVS}\xspace}
\newcommand{\Oh}{\mathcal{O}}
\newcommand{\N}{\mathbb{N}}

\newcommand{\ftfvs}{{$r$-\textit{ftfvs}}\xspace}
\newcommand{\rftfvsp}{{\sc $r$-Fault Tolerant Feedback Vertex Set}\xspace}

\newcommand{\fvs}{fvs\xspace}

\newcommand{\multicutInForests}{\textsc{Multicut in Forests}\xspace}
\newcommand{\multicutInForestsShort}{\textsc{MCF}\xspace}

\usepackage{etoolbox}

\newcommand*\linenomathpatch[1]{%
  \cspreto{#1}{\linenomath}%
  \cspreto{#1*}{\linenomath}%
  \csappto{end#1}{\endlinenomath}%
  \csappto{end#1*}{\endlinenomath}%
}

\linenomathpatch{equation}
\linenomathpatch{gather}
\linenomathpatch{multline}
\linenomathpatch{align}
\linenomathpatch{alignat}
\linenomathpatch{flalign}

\newcommand{\prob}[3]{
  \vspace{2mm}
\noindent\fbox{
  \begin{minipage}{0.96\textwidth}
  \vspace{-1mm}
  \begin{tabular*}{\textwidth}{@{\extracolsep{\fill}}lr} \textsc{#1}
  \vspace{-1mm}
\\ \end{tabular*}
  {\bf{Input:}} #2

  {\bf{Output:}} #3
  \end{minipage}
  }
  \vspace{2mm}
}

\usepackage[textwidth=20mm,obeyFinal,colorinlistoftodos]{todonotes}

\DeclareMathOperator{\dist}{\ensuremath{\operatorname{dist}}}
\DeclareMathOperator{\OPT}{\ensuremath{\operatorname{OPT}}}

\graphicspath{{./images/}}

\journal{Discrete Optimization}

\begin{document}

\begin{frontmatter}

\title{Constant Factor Approximation for Tracking Paths and Fault Tolerant Feedback Vertex Set\tnoteref{tn1}}
\tnotetext[tn1]{A preliminary version of this work is to appear in the proceedings of WAOA 2021. Apart from the full proofs, this version improves the approximation factors for \textsc{$r$-Fault Tolerant Feedback Vertex Set} and \textsc{Multicut in Forests}.\\
The authors acknowledge the support of the OP VVV MEYS funded
project CZ.02.1.01/0.0/0.0/16\_019/0000765 ``Research Center for Informatics''.
This work was supported by the Grant Agency of the Czech Technical University
in Prague, grant \mbox{No.~SGS20/208/OHK3/3T/18}.}


\author{V\'aclav Bla\v zej}
\author{Pratibha Choudhary}
\author{Du{\v s}an Knop}
\author{Jan Maty{\'a}{\v s} K{\v r}i{\v s}{\v t}an\corref{cor1}}
\author{Ond\v rej~Such\'{y}}
\author{Tom{\'a}{\v s} Valla}
\fntext[fn1]{{\it Email addresses:} \tt \{blazeva1, pratibha.choudhary, dusan.knop, kristja6, ondrej.suchy, tomas.valla\}@fit.cvut.cz}

\cortext[cor1]{Corresponding author; kristja6@fit.cvut.cz}

\affiliation{organization={Department of Theoretical Computer Science, Faculty of Information Technology, Czech Technical University in Prague},
            addressline={Thakurova 9},
            city={Prague},
            postcode={160 00},           
            country={Czech Republic}}


\begin{abstract}
Consider a vertex-weighted graph $G$ with a source $s$ and a target $t$.
\tprob requires finding a minimum weight set of vertices (\emph{trackers}) such that the sequence of trackers in each path from $s$ to $t$ is unique.
In this work, we derive a factor $6$-approximation algorithm for \tprob in weighted graphs and a factor $4$-approximation algorithm if the input is unweighted.
This is the first constant factor approximation for this problem.
While doing so, we also study approximation of the closely related \textsc{$r$-Fault Tolerant Feedback Vertex Set} problem.
There, for a fixed integer~$r$ and a given vertex-weighted graph $G$, the task is to find a minimum weight set of vertices intersecting every cycle of~$G$ in at least~$r+1$ vertices.
We give a factor $\mathcal{O}(r)$ approximation algorithm for \textsc{$r$-Fault Tolerant Feedback Vertex Set} if $r$ is a constant.
\end{abstract}



\end{frontmatter}


\section{Introduction}
\label{sec:intro}

In this paper, we study the \tprob problem, which involves finding a minimum weight set of vertices in a vertex-weighted simple graph~$G$ that can track moving objects in a network along the way from a source~$s$ to a target~$t$.
A set of vertices~$T$ is a \emph{tracking set} if for any (simple) \stpath~$P$ in $G$ the sequence $\overrightarrow{T_P}$ of the subset of vertices from~$T$ that appear in $P$, in their order along~$P$, uniquely identifies the path~$P$.
That is, if $\overrightarrow{T_{P_1}}\neq \overrightarrow{T_{P_2}}$ holds for any two distinct \stpath{}s $P_1$ and $P_2$.
Formally, the problem is the following.

\prob{\tprob}
{Undirected graph $G$ with positive integer weights $w \colon V(G) \to \N$, a source~$s$, and a target~$t$.}
{A minimum weight tracking set $T\subseteq V(G)$.}

\paragraph{Motivation for Tracking Paths}
The problem is motivated by the nature of information exchange in complex settings.
In the current age of information, social media networks have an important role in information exchange and dissemination.
However, due to the unregulated nature of this exchange, spreading of rumours and fake news pose serious challenges in terms of authenticity of information~\cite{ChierichettiLP11,rumour-spreading}.
Identifying and studying patterns of rumor spreading in social media poses a lot of challenges due to huge amounts of data in constant movement in large networks~\cite{rumour-det}.
Tracing the sequence of channels (people, agents,~\dots) through which rumors spread can make it easier to contain the spread of such unwanted messages~\cite{rumour-bigdata,rumour-pattern}.
A basic approach would require tracing the complete route traversed by each message in the network.
Here, an optimum tracking set can serve as a resource-efficient solution for tracing the spread of rumors and dissolving them.
Furthermore, \tprob finds applications in tracking traffic movement in transport networks and tracing object movement in wireless sensor networks~\cite{sensor-tracking,localization}.

\paragraph{Known results}
The graph theoretic version of the problem was introduced by Banik et al.~\cite{BanikKPS20}, wherein the authors studied the unweighted (i.e., $w(v) = 1$ for all $v \in V(G)$) shortest path variant of the problem, namely \textsc{Tracking Shortest Paths} (i.e., the set~$T$ is required to uniquely identify each of the shortest \stpath{}s).
They showed that this problem is \textsc{NP}-hard, 
even to approximate it within a factor of $1.1557$.
They also show that \textsc{Tracking Shortest Paths} admits a $2$-approximation algorithm for planar graphs.
Later, the parameterized complexity of \tprob was studied in~\cite{tr-j}, where the problem was also proven to be \textsc{NP}-hard.

To the best of our knowledge, Eppstein et al.~\cite{ep-planar} were the first to study approximation algorithms for the unweighted \tprob.
They gave a $4$-approximation algorithm when the input graph is planar.
Recently, this result was extended by Goodrich et al.~\cite{ep-log} to a $(1+ \varepsilon)$-approximation algorithm for $H$-minor free graphs and an $\Oh(\log\OPT)$-approximation algorithm for general (unweighted) graphs~\cite{ep-log}.
They also gave an $\Oh(\log n)$-approximation algorithm for \tprob.
The existence of constant factor application algorithm was posed as an open problem by Eppstein et al.~\cite{ep-planar}.
In this paper, we answer this affirmatively.

\begin{theorem}\label[theorem]{thm:tracingPaths:constantFactorApproximation}
	There is a $6$-approximation algorithm for \tprob in weighted graphs and a $4$-approximation algorithm if the input is unweighted.
\end{theorem}

There exists an interesting connection between \fvsp (\FVS) and \tprob.
Before we discuss this in more detail, we introduce \FVS and its fault tolerant variant.
Formally, for a given vertex-weighted graph $G=(V,E)$, \FVS requires finding a minimum weight set of vertices $S\subseteq V$, referred to as a \textit{feedback vertex set} (\fvs), such that the graph induced by the vertex set $V\setminus S$ does not have any cycles.
\FVS is a classical \textsc{NP}-hard problem~\cite{Karp72} that has been thoroughly studied in graph theory.
An \emph{$r$-fault tolerant feedback vertex set} (\ftfvs) is a set of vertices that intersect with each cycle in the graph in at least $r+1$ vertices; finding a minimum weight \ftfvs is the \rftfvsp problem~\cite{pm-1fvs}:

\prob{\rftfvsp}{Undirected graph $G$, weight function $w \colon V \to \N$.}{A minimum weight set of vertices $S\subseteq V(G)$ such that for each cycle $C$ in $G$, it holds that $|V(C)\cap S|\geq r+1$.}

Note that if a graph has a cycle of length less than or equal to $r$, then it cannot have an \ftfvs.

\paragraph{Relation Between \FVS and \tprob}
For a graph $G$ with source $s$ and destination $t$, if each vertex and edge participates in an \stpath, then we refer to $G$ as a preprocessed graph.
It is known that in a preprocessed graph, a tracking set is also a feedback vertex set~\cite{tr-j}.
Thus, the weight of a minimum feedback vertex set serves as a lower bound for the weight of a tracking set in preprocessed graphs. This lower bound has proven to be helpful in the analysis of \tprob. However, approximating \tprob has been challenging since the size of a tracking set can be arbitrarily larger than that of a minimum \fvs.

Furthermore, it is known~\cite{iwoca,ep-planar} that in a graph $G$, if a set of vertices $T$ contains at least three vertices from each cycle in $G$, then $T$ is a tracking set for $G$.
Thus, a $2$-fault tolerant feedback vertex set is also a tracking set.
In this paper, we borrow inspiration from this concept to derive a polynomial time algorithm to compute an approximate tracking set. In particular, we start with finding an fvs for the input graph $G$, and then  identify cycles that need more vertices as trackers additional to the ones selected as feedback vertices.

Observe that a feedback vertex set is indeed a $0$-fault tolerant \fvs.
Misra~\cite{pm-1fvs} gave a $3$-approximation algorithm for the problem of finding a $1$-fault tolerant \fvs in unweighted graphs and $34$-approximation algorithm for weighted graphs.
In this paper, we give an approximation algorithm for finding an $r$-fault tolerant feedback vertex set, where $r$ is a constant.
We do this by using the \multicutInForests problem (see \Cref{sec:multicut}) as an auxiliary problem.
Misra~\cite{pm-1fvs} pointed out that the complexity of \rftfvsp is not known for~$r \ge 2$ and asked for an approximation algorithm.

\begin{theorem}\label{thm:approximationForRRedundantFVS}
There is a $2r+2$-approximation algorithm for \rftfvsp in weighted graphs, and a $r+2$-approximation algorithm if the input is unweighted, where $r$ is a constant.
\end{theorem}

It is worth mentioning that our approach relies on the explicit enumeration of certain cycles in the input graph~$G$.
This can be done in polynomial time if $r$ is a constant (see \Cref{obs:higherRedFVS:computePcomplexity}).
Thus, it remains open how to approximate the \rftfvsp if $r$ depends on the size of the input.


\paragraph{Motivation for (Fault Tolerant) \FVS}
The \FVS problem is motivated by applications in deadlock recovery~\cite{GardarinS76,SiberschatzG93}, Bayesian inference~\cite{Bar-YehudaGNR98}, VLSI design~\cite{HudliH94}, and other areas.
Fault tolerant solutions are crucial to real world applications that are prone to failure of nodes in a network or entities in a system~\cite{Parter16}.
In the case of \FVS, the failure corresponds to not being able to eliminate the node from the network.
%

\paragraph{Related work}
There has been a lot of heuristic-based work on the problem of tracking moving objects in a network~\cite{network-tracking,tracking-info,ZhouM19}.
Parameterized complexity of \textsc{Tracking Shortest Paths} and \textsc{Tracking Paths} was studied in~\cite{tr-j,tr1-j,BiloGLP20,quad,struct-tp,ep-planar}.
\fvsp is known to admit a $2$-approximation algorithm which is tight under UGC~\cite{BafnaBF99,ChudakGHW98}.
The best known parameterized algorithm for \FVS runs in $2.7^k \cdot n^{\Oh(1)}$ time, where $k$ is the size of the solution~\cite{LiN20}.
It is worth noting that Misra~\cite{pm-1fvs} uses \multicutInForests as a subroutine as well.
The edge version is known to admit an LP formulation whose matrix is totally unimodular~\cite{GolovinNS06} if the family of paths is non-crossing; thus, it is solvable in polynomial time.
A related problem is the \textsc{$d$-Hurdle Multiway Cut} for which a factor $2$-approximation algorithm is known (and it is again tight under UGC)~\cite{DeanGPW11}.

\paragraph{Preliminaries and Notations}
We refer to~\cite{diestel} for the standard graph theory terminology.
All paths that we consider in this work are simple paths.
For a graph $G$, we use $V(G)$ to denote its vertex set.
For a set $S \subseteq V(G)$ we use $G\setminus S$ to denote a graph that results from the deletion of the vertex set $S$ and the edges incident to $S$.
For a weight function $w \colon V(G) \rightarrow \mathbb{N}$, let $w(S)$ for $S \subseteq V(G)$ denote the sum of respective elements, i.e., $\sum_{v\in S}w(v)$.
An unweighted version of the problem is obtained by assigning all vertices the same weight.
In fact, we can also omit the weights in this case.
We write vectors in boldface and their entries in normal font, i.e., $x_3$ is the third entry of a vector~$\bm{x}$.
If we apply minimum to two vectors, then it is applied entry-wise.
We write $\bm{1}^n$ for the vector of $n$ ones; we omit the superscript if the dimension is clear from the context.

%

\section{Vertex Multicut in Forests}
\label{sec:multicut}

In this section, we gather polynomial time (approximation) algorithms for solving \textsc{Vertex Multicut in Forests}. The algorithms are used later in the subsequent sections to derive the approximation algorithms for \textsc{$r$-Fault Tolerant \FVS} and \tprob.

\prob{\multicutInForests (\multicutInForestsShort)}
{A forest $F = (V,E)$, weight function $w \colon V \to \N$, terminal pairs $(s_1, t_1), \ldots, (s_\ell, t_\ell)$.}
{A minimum weight set of vertices $S\subseteq V$ such that in the graph~$F~\setminus~S$ there is no path between $s_i$ and~$t_i$ for $i\in[\ell]$.}

In this work, we consider the \textsc{Unrestricted} version of \multicutInForests i.e., a solution set is allowed to contain vertices from the terminal pairs.
We will consider a set of paths $\mathcal{P}$ instead of a set of terminal pairs.
It is not hard to see that these two versions are equivalent on forests: if the terminals in a pair belong to different trees, then we can discard the pair since there does not exist any path between them, otherwise there exists a unique path between each pair of terminals, which the solution~$S$ should intersect.

While \multicutInForestsShort{} is NP-hard~\cite{CalinescuFR03}, the unweighted version is polynomial time solvable \cite{CalinescuFR03,GuoHKNU08}.
Recently Agrawal et al.~\cite[Lemma 5.1]{AgrawalLMSZ20} showed LP relative constant factor approximability of the problem for all chordal graphs. Next we simplify their proof and adapt it to forests, improving the approximation factor from 32 to 2.

Consider the natural ILP formulation of the problem, where we introduce a binary variable $x_v \in \{0,1\}$ for each vertex $v \in V$, describing whether the corresponding vertex should or should not be taken into a solution. At least one vertex must be taken from each path from $\mathcal{P}$ to construct a feasible solution.
Consider the following LP relaxation of the natural ILP.

\begin{equation}\tag{LP$_{\operatorname{MCF}}$}\label{lp:multicutInForests}
\begin{aligned}
& \text{Minimize} & \sum_{v \in V} w(v)\cdot x_v & & \\
& \text{subject to} & \sum_{v \in V(P)} x_v &\ge 1 &&&\forall P\in\mathcal{P} \\
& &  0\le x_v &\le 1  &&&\forall v\in V.
\end{aligned}
\end{equation}

\begin{lemma}\label[lemma]{lem:weighted_multicut}
    For a given instance of \textsc{Multicut in Forests} one can find a solution $S$ such that $w(S) \le 2 \cdot \OPT$ in polynomial time, where $\OPT$ is the objective value of an optimal solution to the corresponding \eqref{lp:multicutInForests}.
\end{lemma}
\DeclarePairedDelimiter{\fractionalPart}{\langle}{\rangle}
\begin{proof}
    Consider an optimal fractional solution $\bm{x}$ to \eqref{lp:multicutInForests} of the \multicutInForestsShort instance (i.e., $w(\bm{x}) = \OPT$).
    Let $c \ge 1$ be a constant (whose value we determine later).
    Let $\widehat{\bm{x}} = \min(c \cdot \bm{x},\bm{1})$, i.e., $\widehat{x}_v = \min(c \cdot x_v, 1)$ for every $v \in V(G)$.

    Without loss of generality we can assume that $G$ is connected as otherwise we can process one connected component at a time.
    Root the graph in an arbitrary vertex $\rho_G$.
    By $P(u,v)$, where $u,v \in V(G)$, we denote the unique path from $u$ to $v$ in $G$.
    Let $d_v = \sum_{u\in P(\rho_G,v)}\widehat{x}_u$ be the distance of $v$ from $\rho_G$ in $G$.
    For a real $a$ define the \emph{fractional part} of $a$ as $\fractionalPart{a} = a - \lfloor a \rfloor$ (e.g., $\fractionalPart{5.9876} = 0.9876$, and $\fractionalPart{-2.345} = 0.655$).

    For $r \in [0,1)$ let us define the \emph{bin of $r$}, denoted $B_r$, as the set of vertices $u \in V(G)$ for which
    \begin{equation}\label{eq:split_vertex_interval}
        r \in \begin{cases}
            \big[\fractionalPart{d_u - \widehat{x}_u}, \fractionalPart{d_u}\big) &\text{if $\fractionalPart{d_u} - \widehat{x}_u \ge 0$,}\\
            \big[\fractionalPart{d_u - \widehat{x}_u}, 1\big) \cup \big[0, \fractionalPart{d_u}\big) &\text{otherwise.}\\
        \end{cases}
    \end{equation}
    Note that a vertex is in $\widehat{x}_u$ proportion of all the bins, and so a vertex with $\widehat{x}_u \ge 1$ belongs to all bins.
    We call intervals $[\fractionalPart{d_u - \widehat{x}_u}, \fractionalPart{d_u})$ \emph{aligned to} $[\fractionalPart{d_v - \widehat{x}_v}, \fractionalPart{d_v})$ if $\fractionalPart{d_v} = \fractionalPart{d_u - \widehat{x}_u}$.
    Similarly for intervals $[\fractionalPart{d_u - \widehat{x}_u}, 1\big) \cup \big[0, \fractionalPart{d_u})$ etc.
    Note that if a vertex $v$ is a predecessor of a vertex $u$ (on the path from $\rho_G$ to $u$), i.e., $v$ is a parent of $u$, then their intervals are aligned.
    See Fig.~\ref{fig:few_different_bins} for an example of ranges of bins which contain vertices and how parent-child pairs align.

    \begin{figure}[h]
        \centering
        \includegraphics[scale=1.2]{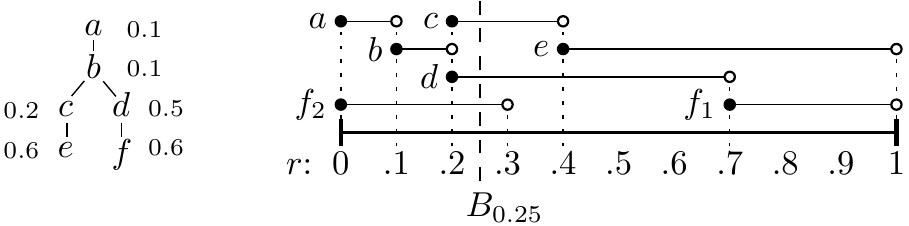}
        \caption{
            Left: A tree $G$ consisting of vertices $a$ through $f$ with $\rho_G=a$ together with their weights $w$.
            Right: The $[0,1)$ range together with intervals of bins which contain respective vertices of $G$.
            E.g., bin $B_{0.25} = \{c,d,f\}$.
            Intervals have closed and open endpoints which are shown as full and empty circles, respectively.
            The values $r$ such that $f \in B_r$ make up two intervals $f_1$ and $f_2$ because the bins were chosen according to the second case of Eq.~\ref{eq:split_vertex_interval}.
        }%
        \label{fig:few_different_bins}
    \end{figure}

    \begin{Claim}
        For every $r \in [0,1)$ we claim that $B_r$ is a solution to the \multicutInForestsShort instance.
    \end{Claim}
    \begin{proof}
        We show this claim by a contradiction.
        Assume that vertices in some $B_r$ do not constitute a solution.
        Then there is a path $P \in \mathcal{P}$ with endpoints $s$ and $t$ which was not disconnected by removal of $B_r$.
        Take $z$ as the last common vertex of paths $P(\rho_G,s)$ and $P(\rho_G,t)$.
        On the one hand, if $\sum_{v\in P(z,s)}\widehat{x}_v \ge 1$, then by aligning the intervals of vertices of $P(z,s)$ (which is possible because each pair of consecutive vertices is in a parent-child relationship) we get an interval which spans range of size at least $1$.
        In such case $B_r$ contains some vertex of $P(z,s)$, a contradiction.
        The same argument also works for $P(z,t)$.
        On the other hand, if both $\sum_{v\in P(z,s)}\widehat{x}_v < 1$ and $\sum_{v\in P(z,t)}\widehat{x}_v < 1$, then we have that for $c=2$
        \begin{align*}
            \sum_{v\in V(P)}\widehat{x}_v &< 2, \\
            \smashoperator{\sum_{v\in V(P)}}\,c\cdot x_v &< 2, \\
            \sum_{v\in V(P)} x_v &< 1.
        \end{align*}
        This is a contradiction with $\bm{x}$ being a solution to \eqref{lp:multicutInForests}, which proves the claim.
    \end{proof}

    \begin{Claim}
        There exists a bin $B_{r^*}$ with $w(B_{r^*}) \le c \cdot \OPT$.
    \end{Claim}
    \begin{proof}
        Choosing $r$ from interval $[0,1)$ uniformly at random we get a bin $B_r$ where the probability that a vertex $v \in V(G)$ belongs to~$B_r$ is $\widehat{x}_v$ because the bins which contain $v$ span exactly $\widehat{x}_v$ out of interval $[0,1)$.
        By the linearity of expectation, we have that the expected weight of elements in $B_r$ is
        \[
            \mathbb{E}\bigg[\sum_{v\in V(G)}w(v) \cdot \mathrm{Ind}[v \in B_r]\bigg]\
            = \smashoperator{\sum_{v\in V(G)}} w(v) \cdot \Pr[v \in B_r]\
            = \smashoperator{\sum_{v\in V(G)}} w(v) \cdot \widehat{x}_v,
        \]
        where $\mathrm{Ind}$ denotes an indicator variable of an event.
        Hence, there exists a bin $B_{r^*}$ with weight
        \[
            w(B_{r^*}) \le \smashoperator{\sum_{v \in V(G)}} w(v) \cdot \widehat{x}_v = w(\widehat{\bm{x}}) \le c \cdot w(\bm{x}) = c \cdot \OPT,
        \]
        which proves the claim.
    \end{proof}

    There exists a value $r^* \in [0,1)$ such that $w(B_{r^*}) \le c \cdot \OPT$, therefore $B_{r^*}$ constitutes a $c$-approximate solution to the \textsc{Multicut in Forest} instance.

    We compute $\bm{x}$ in polynomial time using the Ellipsoid method.
    Then, it is elementary to compute $\widehat{x}_v$ and $d_v$ for every $v \in V(G)$ according to their definitions.
    If we consider the $[0,1)$ range as a continuous looped segment (by identifying $0$ with~$1$) we see that each vertex is contained in a continuous interval of bins.
    Consider adding such intervals one by one to an initially empty range.
    The range splits into continuous \emph{bin-groups} that contain the same set of vertices.
    Each added interval may split up to two bin-groups into two.
    Hence, there are at most $2 \cdot |V(G)|$ bin-groups, see Fig.~\ref{fig:few_different_bins}.
    By going through each interval endpoint, we can go through all bin-groups.
    We can now go through each interval endpoint (and zero) and evaluate the sum over weights of vertices in the bin-group.

    By taking the smallest-weighted bin-group, we get a bin which is as good as $B_{r^*}$ -- it is a $c$-approximate solution to the \textsc{Multicut in Forest} instance, which concludes the proof of the lemma.
\end{proof}

The following observation shows that the improvement of the approximation factor below 2 is unlikely.

\begin{observation}
 There is an approximation preserving reduction from \textsc{Vertex Cover} to \textsc{Multicut in Forest}.
\end{observation}

\begin{proof}
 Let $G=(V,E)$ and $w \colon V \to \N$ form an instance of weighted \textsc{Vertex Cover}.
 The desired instance of \textsc{Multicut in Forest} is formed by a star with $V$ being the set of leaves.
 The weight of leaves is retained, while we assign some heavy weight, such as $\sum_{v\in V} w(v)$, to the centre $c$ of the star, so that it is never selected in a solution.
 The paths to be cut are formed by the edges in $E$, that is, for each edge $\{u,v\} \in E$ we introduce the path $u,c,v$ to $\mathcal{P}$.
 It is easy to observe that a subset of $V$ is solution to the formed instance of \textsc{Multicut in Forest} if and only if it is a vertex cover in $G$. As the weights are the same, the reduction preserves the inapproximability.
\end{proof}

\subsection{Chordal Graphs}

It is worth mentioning that the simplifications for the special case of trees might be transfered to the case of chordal graphs.
Therefore, we improve the approximation factor of Agrawal et al.~\cite[Lemma 5.1]{AgrawalLMSZ20} for general chordal graphs from 32 to 4.
We believe that this is of independent interest.

Note that in chordal graphs there might be several paths between each terminal pair, hence, we have to stick to terminal pairs.
The following constitutes an analogue of \Cref{lem:weighted_multicut} for chordal graphs.

\begin{figure}[h]
    \centering
    \caption{The grey rectangles indicate vertices of the clique tree $T$, i.e. cliques of $G$. The grey edges indicate paths in $T$, as the shown vertices of $T$ need not be directly connected. The root of $T$ is on the left, with depth of the tree increasing to the right. Note that in general one vertex of $G$ can be contained in multiple vertices of $T$. Also $z$, $u$, $s'$ and $t'$ need not be distinct vertices.}
    \includegraphics[scale=1.2,page=3]{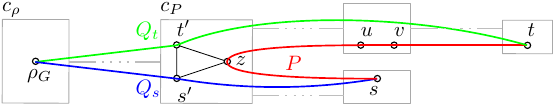}
    \label{fig:chordal_graph}
\end{figure}

\begin{lemma}
    For a given instance of \textsc{Multicut in Chordal Graphs} one can find a solution $S$ such that $w(S) \le 4 \cdot \OPT$ in polynomial time, where $\OPT$ is the objective value of an optimal solution to the corresponding LP. 
\end{lemma}
\begin{proof}
    Consider again an optimal fractional solution $\bm{x}$ to the LP of the \textsc{Multicut in Chordal Graphs} instance (i.e., $w(\bm{x}) = \OPT$).
    Let $c = 4$ and $\widehat{\bm{x}} = c \cdot \bm{x}$.
    Our aim is to show that there is an integral solution of weight $w(\widehat{\bm{x}})$.
    To this end, we can afford to take all vertices $v$ with $\widehat{x}_v \ge 1$ into the solution and remove them from the graph.
    Hence, we assume that for each vertex $v$ remaining in the graph we have $\widehat{x}_v < 1$.

    Without loss of generality we assume that $G$ is connected.
    Root the graph in an arbitrary vertex $\rho_G$.
    For a path $P$ in $G$ let the length $\ell(P)$ of $P$ be defined as $\sum_{u\in P}\widehat{x}_u$.
    For each $v \in V(G)$ we let $d_v$ be the length of a shortest path from $\rho_G$ to $v$, i.e., minimum of $\ell(P)$ over all paths $P$ from $\rho_G$ to $v$. 
    
    With this definition of $d_v$ we define the bins exactly as in \Cref{lem:weighted_multicut}.
    Note that $v \in B_r$ if and only if there is an integer $q$ such that $d_v -x_v \le q+r < d_v$ (this integer can only be $q= \lceil d_v-x_v-r \rceil$, since $x_v <1$).
    \begin{Claim}
        For every $r \in [0,1)$ we claim that $B_r$ is a solution to the \textsc{Multicut in Chordal Graphs} instance.
    \end{Claim}
    \begin{proof}
        We show this claim by a contradiction.
        Assume that vertices in some $B_r$ do not constitute a solution.
        Then, there is a pair $(s,t)$ of terminals which were not disconnected by removal of $B_r$, i.e., there is a path $P$ from $s$ to $t$ in $G \setminus B_r$.
        Consider a clique tree $T$ of $G$.
        Let $c_\rho$ be an arbitrary clique containing $\rho_G$ and root $T$ in $c_\rho$.
        Let $c_P$ be a clique such that $c_P$ contains a vertex of $P$ and all vertices of $P$ are only contained in descendants of $c_P$.
        Such a clique exists due to the properties of the clique tree.
        Note that for each vertex $u$ of $P$, each path from $\rho_G$ to $u$ contains a vertex of $c_P$.
        See Fig. \ref{fig:chordal_graph} for an illustration.
        
        Let $z$ be an arbitrary vertex from $P$ in $c_P$.
        Let $q$ be the maximum integer such that $q+r-1 < d_z -x_z$, i.e., $q+r \ge d_z -x_z$ and $q= \lceil d_z-x_z-r \rceil$.
        Suppose first that there is a vertex $v$ on $P$ such that $d_v  > q+r$.
        If this is the case for $z$, then we have $d_z -x_z \le q+r < d_z$ and $z \in B_r$, contradicting the choice of $P$.
        Thus suppose that $v$ is on the part of $P$ from $z$ to $t$, the other case is symmetric.
        Let $v$ be the first such vertex on that path and $u$ be the predecessor of $v$ on this part of $P$.
        Then we have $d_u \le q+r$ and by triangle inequality $d_v \le d_u+x_v$, implying $d_v-x_v \le d_u \le q+r$.
        But then $d_v -x_v \le q+r < d_v$ and $v$ is in $B_r$, contradicting the choice of $P$.
        
        Therefore, for each vertex $v$ in $P$ we have $d_v \le q+r$ and, hence, $d_v -(d_z-x_z) < q+r -(q+r-1) =1$.
        In particular, $d_s- (d_z-x_z) <1$ and $d_t- (d_z-x_z) <1$.
        Consider a path $Q_s$ from $\rho_G$ to $s$ of length $d_s$.
        This path must contain a vertex of $c_P$, let us denote an arbitrary such vertex $s'$.
        Similarly, let $t'$ be a vertex of $c_P$ on a path $Q_t$ of length $d_t$ from $\rho_G$ to $t$.
        As $s'$ and $z$ are both in clique $c_P$, they are adjacent and by triangle inequality we have $d_z \le d_{s'}+x_z$, i.e., $d_z -x_z \le d_{s'}$.
        Let $Q'_s$ be the part of $Q_s$ from $s'$ to $s$.
        We have $\ell(Q'_s) = d_s - d_{s'}+ x_{s'} \le d_s- (d_z-x_z) + x_{s'} < 1+1 =2$.
        Similarly, the part $Q'_t$ of $Q_t$ from $t'$ to $t$ has length less than $2$.
        As both $s'$ and $t'$ are contained in $c_P$, they are adjacent and we can join $Q'_s$ and $Q'_t$ into one path  $Q'$ from $s$ to $t$ of total length less than $4$. 
        Then we have $4 > \sum_{u\in Q'}\widehat{x}_u = \sum_{u\in Q'}c \cdot x_u$. 
        Since $c=4$, we get that $\sum_{u\in Q'} x_u <1$ contradicting $\bm{x}$ being a feasible solution to the LP of the instance, which proves the claim.
    \end{proof}
  The rest of the proof is exactly the same as in the proof of \Cref{lem:weighted_multicut}.
\end{proof}

\subsection{Unweighted Version}
We also mention the following result of similar nature for unweighted forests, strengthening the polynomial time solvability of the problem.

\begin{lemma}\label[lemma]{lem:optimal_integral}
If all the weights are equal, then there is an integral optimal solution for \eqref{lp:multicutInForests}.
Furthermore, such a solution can be found in polynomial time.
\end{lemma}
\begin{proof}
It is enough to show the lemma for each tree $T$ of $F$ separately.

Root the tree $T$ in an arbitrary vertex $r$.
Among all optimal solutions to the LP, consider the solution~$\bm{x}$ that minimizes $\sum_{v \in V(T)} x_v \cdot \dist(v, r)$.

  Assume for the sake of contradiction, that~$\bm{x}$ is not integral.
  Let $u$ be a vertex with $0 < x_u < 1$ and the maximum distance from the root~$r$.
Note that, in particular, for all descendants~$v$ of~$u$, it holds that $x_v \in \{ 0,1 \}$.
Let us first assume that $u \neq r$ and let $p$ be the parent of~$u$.
We define $\bm{\widehat{x}}$ as
\[
  \widehat{x}_v =
  \begin{cases}
    0 &\text{if } v = u, \\
    \min\{x_p+x_u,1\} &\text{if } v = p, \\
    x_v & \text{otherwise.}
  \end{cases}
\]

We claim that $\bm{\widehat{x}}$ is a solution to \eqref{lp:multicutInForests}.
Obviously, $0\le \widehat{x}_v \le 1$ for every $v\in V$.
Let $P \in \mathcal{P}$.
If $u \notin V(P)$, then $\sum_{v \in V(P)} \widehat{x}_v \ge \sum_{v \in V(P)} x_v \ge 1$.
If $u \in V(P)$, then either the path $P$ is fully contained in the subtree of $F$ rooted in $u$, or $p \in  V(P)$.
In the first case, we have $\sum_{v \in V(P)} \widehat{x}_v \ge \big(\sum_{v \in V(P)} x_v\big)  - x_u\ge 1-x_u > 0$.
Since the summands on the left hand side are integral, by the choice of $u$, it follows that $\sum_{v \in V(P)} \widehat{x}_v \ge 1$.
In the later case, if $\widehat{x}_p=1$, then $\sum_{v \in V(P)} \widehat{x}_v \ge 1$ and otherwise  $\sum_{v \in V(P)} \widehat{x}_v =\big(\sum_{v \in V(P)} x_v\big) +x_u - x_u=\sum_{v \in V(P)} x_v\ge 1$.

Hence, $\bm{\widehat{x}}$ is a solution to \eqref{lp:multicutInForests}.
Furthermore, we have that $\sum_{v \in V} w(v) \cdot\widehat{x}_v \le \sum_{v \in V} {w(v) \cdot x_v}$ (since all weights are equal) and
\(
  \sum_{v \in V} x_v \cdot \dist(v, r) > \sum_{v \in V} \widehat{x}_v \cdot \dist(v, r)
\)
which is a contradiction.
The case $u=r$ can be proved using a similar argument for $\bm{\widehat{x}}$ such that $\widehat{x}_u = 0$ and $\widehat{x}_v=x_v$ if $v \neq u$.

The second part of the lemma follows from polynomial time algorithm for \textsc{MCF}~\cite{CalinescuFR03}, since any optimal solution to the instance of \textsc{MCF} represents an optimal solution for the corresponding \eqref{lp:multicutInForests} by the first part of the lemma.
\end{proof}

For the rest of the paper, we use $\mu$ to denote the best approximation ratio achievable for \textsc{MCF} with respect to LP relative approximation.
That is, $\mu \le 2$ for weighted instances and $\mu=1$ for unweighted instances.

\input{r-ftfvs-approximation}

\section{Approximate Tracking Set}

In this section, we give a constant factor approximation algorithm for \tprob.

Let $G=(V,E)$ be the input graph and $s$ and $t$ the source and the target.
We start by applying the following reduction rule on~$G$.
This can be done in polynomial time~\cite{tr-j}.
%

\begin{Reduction Rule}[Banik et al.~\cite{tr-j}]
\label[Reduction Rule]{red:stpath}
If there exists a vertex or an edge that does not participate in any \stpath, then delete it.
\end{Reduction Rule}

We use the term \textit{preprocessed graph} to denote a graph resulting from exhaustively applying~\ref{red:stpath}.
For the sake of simplicity, after the application of reduction rule, we continue to refer to the reduced graph as $G$.

Next, we describe the \emph{local source-destination pair (local  $s$-$t$ pair)}, a concept that has served as crucial for developing efficient algorithms for \tprob~\cite{tr-j,iwoca,struct-tp,ep-planar}. For a subgraph $G'\subseteq G$, and vertices $a,b\in V(G')$, we say that $a,b$ is a \emph{local  $s$-$t$ pair} for $G'$ if
\begin{enumerate}
\item there exists a path in $G$ from $s$ to $a$, say $P_{sa}$,
\item there exists a path in $G$ from $b$ to $t$, say $P_{bt}$,
\item $V(P_{sa})\cap V(P_{bt})=\emptyset$, and
\item $V(P_{sa})\cap V(G')=\{a\}$ and $V(P_{bt})\cap V(G')=\{b\}$.
\end{enumerate}
Note that a subgraph can have more than one local source-destination pair.

\begin{lemma}\label[observation]{obs:localst}
Let $G$ be a graph and let $G'$ be a subgraph of $G$.
We can verify in $\mathcal{O}(n^2)$ time whether $a,b \in V(G')$ is a local $s$-$t$ pair for $G'$.
\end{lemma}
\begin{proof}
A pair of vertices $a,b\in V(G')$ forms a local source-destination pair for $G'$ if and only if there exist disjoint paths from $s$ to $a$ and $b$ to $t$ in the graph $G\setminus (V(G')\setminus \{a,b\})$.
It can be verified in $\mathcal{O}(n^2)$ time whether $a,b$ satisfy this condition using the disjoint path algorithm from~\cite{KAWARABAYASHI2012424}.
\end{proof}

We recall the following lemma from the previous work.

\begin{lemma}[{\cite[Lemma~2]{iwoca}}]\label[lemma]{lem:notTrackingImpliesCycleAndSTPair}
  In a graph~$G$, if $T \subseteq V(G)$ is not a tracking set for~$G$, then there exist two \stpaths with the same sequence of trackers, and they form a cycle~$C$ in~$G$, such that $C$ has a local source~$a$ and a local destination~$b$, and $T \cap (V(C) \setminus \{ a,b \}) = \emptyset$.
\end{lemma}


Eppstein et al.~\cite{ep-planar} mentioned that a $2$-fault tolerant feedback vertex set is always a tracking set. Here we use a variation of this idea to compute an approximate tracking set. Specifically, we start with a $2$-approximate feedback vertex set and then identify the cycles that contain only one or two feedback vertices. We check if these cycles need more vertices as trackers and we use \eqref{lp:higherRedFVS:1} and \eqref{lp:multicutInForests} explained in the previous section to add them.

Now we present the algorithm for computing a $2(1+\mu)$-approximate tracking set in polynomial time.
We start by computing a $2$-approximate feedback vertex set $S$ on the preprocessed graph $G$ using the algorithm by Bafna et al.~\cite{BafnaBF99}.
We first check whether $S$ is a tracking set for $G$ by using the tracking set verification algorithm given in~\cite{tr-j}.
If it is a tracking set, we return $S$ as the solution, otherwise we proceed further.

If $S$ is not a tracking set, we will find vertices on which to place additional trackers in the following way.
First we identify cycles $C$ such that $|V(C)\cap S|=1$.
Each such cycle $C$ can be obtained by taking a vertex $a \in S$ together with a path between a pair of its neighbors in $G\setminus S$.
For each vertex $b \in V(C) \setminus \{a\}$ we check whether $a,b$ (or $b,a$) is a \lstpair for $C$.
If this is the case, then we do the following.
If $C \setminus \{a, b\}$ has two connected components $P_1, P_2$, then we add the pair $\{P_1, P_2\}$ to $\mathcal{P}$.
Otherwise, if $P_1$ is the only connected component of $C \setminus \{a, b\}$, we add the set $\{P_1\}$ to $\mathcal{P}$.
Note that $P_1$ and $P_2$ will always be paths.

If a cycle $C$ in $G$ intersects with $S$ in vertices $a$ and $b$, then there exist two vertex-disjoint paths $P'_1$ and $P'_2$ between $a$ and $b$, such that $V(P'_1)\cup V(P'_2)=V(C)$.
Hence, each such cycle $C$ is uniquely determined by the neighbors of $a$ and $b$ on $P'_1$ and $P'_2$.
If, furthermore, $a,b$ (or $b,a$) is a \lstpair for $C$, then we add some pair to $\mathcal{P}$.
Similarly to the above, we consider the connected components of $C \setminus \{a, b\}$. If there are two connected components, say $P_1, P_2$, we add $\{P_1, P_2\}$ to $\mathcal{P}$.
Otherwise, $P_1$ is the only connected component of $C \setminus \{a, b\}$ and we add the set $\{P_1\}$ to $\mathcal{P}$.

Similarly to \Cref{obs:higherRedFVS:computePcomplexity}, we have at most $n^2$ candidate cycles with a single vertex of $S$ and for each of them we have at most $n$ candidates on $b$. We have $n^4$ cycles with two vertices of $S$. For each of them, we check, whether $a,b$ (or $b,a$) is a \lstpair in $\Oh(n^2)$ time. Hence, $\mathcal{P}$ can be obtained in $\Oh(n^6)$ time.

Now we use \eqref{lp:higherRedFVS:1} with $\mathcal{P}$ to identify the paths on which we want to place at least one additional tracker.
Let $\bm{x^*}$ be an optimal solution of \eqref{lp:higherRedFVS:1}, which can be obtained in polynomial time using the Ellipsoid method.
We construct $\mathcal{P}_{x^*}$ as the set of all paths $P \in \mathcal{P}$ such that $\sum_{v \in V(P)} x^*_v \ge \frac{1}{2}$.

We first show the following observation.

\begin{observation}\label[observation]{obs:tsXBound}
  Let $G$ be a preprocessed graph and $T^*$ be a tracking set for $G$. Let $\OPT_{x^*}$ be the optimum value of \eqref{lp:higherRedFVS:1}.
  Then $\OPT_{x^*} \leq w(T^*)$.
\end{observation}
\begin{proof}

	Let $\bm{\widehat{x}}$ be a vector such that for all $v \in V(G)\setminus S$
  \[
    \widehat{x}_v=
    \begin{cases}
      1 &\text{if } v \in T^* \\
      0 & \text{otherwise.}
    \end{cases}
  \]
  We show that $\bm{\widehat{x}}$ is a solution to \eqref{lp:higherRedFVS:1}.
  Suppose that $\sum_{v \in V(P_i) \cup V(P_j)} x_v < 1$ for some ${\{P_i, P_j\} \in \mathcal{P}}$, therefore $T^* \cap (V(P_i) \cup V(P_j)) = \emptyset$.
  Let $a, b$ be the vertices such that $G[V(P_i) \cup V(P_j) \cup \{a, b\}]$ contains a cycle $C$, such that $a,b$ are a \lstpair for $C$.
  Such $a,b$ must exist because of the way~$\mathcal{P}$ was constructed.

  Since $a,b$ is a \lstpair, there exist two distinct paths $\widehat{P}_1$ and $\widehat{P}_2$ such that $\overrightarrow{T^*_{\widehat{P}_1}}=\overrightarrow{T^*_{\widehat{P}_2}}$: Both reach $a$ from $s$ and reach $t$ from $b$ but one reaches $b$ from $a$ by $P_i$ and the other one by $P_j$.
	This contradicts the assumption that $T^*$ is a tracking set.
\end{proof}

To decide which additional vertices to include in the solution, we compute the minimum multicut in forests on $(G \setminus S,w|_{(V\setminus S)}, \mathcal{P}_{x^*})$ using \eqref{lp:multicutInForests}.
Let $\bm{y^*}$ be an integral solution of \eqref{lp:multicutInForests} of weight at most $\mu$ times the optimal one.
By \Cref{lem:weighted_multicut} or \Cref{lem:optimal_integral} such a solution can be obtained in polynomial time.
Let $X$ be the set of vertices such that $y_v = 1$.
Then we claim that $S \cup X$ is a $2(1+\mu)$-approximate solution to the \tprob.

\begin{lemma}
  Let $S$ be a $2$-approximate \fvs in~$G$ and let $\bm{y}$ be an integral solution to \eqref{lp:multicutInForests} induced by~$\mathcal{P}_{\bm{x}^*}$ with weight at most $\mu$ times the optimal.
  The set $T = S \cup \left\{ v \in V \setminus S \mid y_v = 1 \right\}$ is a $2(1+\mu)$-approximate solution to \tprob on~$G$.
\end{lemma}
\begin{proof}
Let $X = \left\{ v \in V \setminus S \mid y_v = 1 \right\}$.
First, we show that $T=S \cup X$ is a tracking set.
Suppose there exists a pair of distinct \stpaths $P_1, P_2$ in $G$ that are not distinguished by $T$, i.e., $\overrightarrow{T_{P_1}}=\overrightarrow{T_{P_2}}$.
	Then, by \Cref{lem:notTrackingImpliesCycleAndSTPair}, there exists a cycle $C$ such that $V(C) \subseteq V(P_1) \cup V(P_2)$ and a \lstpair $a,b\in V(C)$ such that $T\cap V(C)\setminus\{a,b\}=\emptyset$.
  However, then $C$ must have been one of the cycles enumerated by the algorithm when constructing $\mathcal{P}$.

  Say $\{P_i, P_j\}$ was the pair added to $\mathcal{P}$ such that $V(P_i) \cup V(P_j) \subseteq V(C)$.
  As $\bm{x}^*$ is the optimal solution to \eqref{lp:higherRedFVS:1}, it must hold $\sum_{v \in V(P_i) \cup V(P_j)} \bm{x}^*_v \geq 1$
  and hence $\sum_{v \in V(P_i)} \bm{x}^*_v \geq \frac{1}{2}$ or $\sum_{v \in V(P_j)} \bm{x}^*_v \geq \frac{1}{2}$.
  Thus $P_i \in \mathcal{P}_{\bm{x}^*}$ or $P_j \in \mathcal{P}_{\bm{x}^*}$.
  In either case, $\sum_{v \in V(P_i) \cup V(P_j)} \bm{y}_v \geq 1$ and therefore there is some $v \in V(P_i) \cup V(P_j)$ which was added to $T$ and distinguishes $\overrightarrow{T_{P_1}}=\overrightarrow{T_{P_2}}$, which contradicts the choice of $C$.


	Now we show that $w(T)$ is at most $2(1+\mu)$ times the weight of a minimum tracking set in~$G$.
	Let $t^*$ be the weight of the minimum tracking set on $G$ and $f^*$ be the weight of a minimum feedback vertex set in $G$.
  Furthermore, let $\OPT_{x^*}$ be the objective value of an optimal solution to \eqref{lp:higherRedFVS:1} and $\OPT_{y^*}$ be the objective value of a solution to \eqref{lp:multicutInForests} induced by $\mathcal{P}_{\bm{x}^*}$.

	We claim that $|S| \leq 2 f^* \leq 2 t^*$. The first inequality follows from the fact that $S$ is a $2$-approximate feedback vertex set for $G$. The second inequality follows from the fact that every tracking set is also a feedback vertex set~\cite{tr-j}.
  It also holds $w(X) \le \mu \OPT_{y^*} \leq 2\mu \cdot\OPT_{x^*} \leq 2\mu \cdot t^*$.
  Here the second inequality follows from \Cref{lem:solutionXBoundOnY}, while the third one follows from~Observation~\ref{obs:tsXBound}.

  Together this gives us $w(T) = w(S) + w(X) \leq 2 t^* + 2 \mu \cdot t^* = 2(1+\mu) \cdot t^*$.
\end{proof}

The result is summed up as follows.

\begin{theorem}[precise version of \Cref{thm:tracingPaths:constantFactorApproximation}]
	There exists a $2(1+\mu)$-approximation algorithm for \tprob, in particular there is a $4$-approximation algorithm for unweighted graphs and $6$-approximation algorithm for weighted graphs.
\end{theorem}

\bibliography{ref}
\end{document}

%% file: r-ftfvs-approximation.tex
\section{Approximate \textsc{$r$-Fault Tolerant Feedback Vertex Set}}

In this section, we give an algorithm for computing an approximate $r$-fault tolerant feedback vertex set for undirected weighted graphs, for any fixed integer $r\geq 1$.
Recall that an $r$-fault tolerant feedback vertex set is a set of vertices that contains at least $r+1$ vertices from each cycle in the graph.
A polynomial time algorithm for computing a constant factor approximate $1$-fault tolerant feedback vertex set was given by Misra~\cite{pm-1fvs}.
The factor can be easily observed to be $2+\mu$, where $\mu$ is the best possible approximation ratio for \textsc{MCF}.
We show a different approach that for any fixed $r \geq 1$ gives an $(2 + r\mu)$-approximate solution.

\input{r-ftfvs-hardness}

\subsection{Computation of approximate $r$-Fault Tolerant Feedback Vertex Set}

For the rest of the section, we assume that $r\geq 1$ is a fixed constant.
Let $G = (V, E)$ be the input graph.
First, we assume that we have a $\alpha$-approximate feedback vertex set $S$, for example, by the algorithm by Bafna et al.~\cite{BafnaBF99}.
Our goal is to compute a vertex set that contains at least $r+1$ vertices from each cycle in $G$.

To achieve that, we first create a family $\mathcal{P}$ of path sets which will eventually be used to select vertices which will be added to $S$ to extend it to an $r$-ftfvs.

To construct $\mathcal{P}$, we consider every subset $X$ of $S$ such that $1 \leq |X| \leq r$.
Each such subset can be a part of many different cycles, we want to consider each such cycle.
To do that, we take each of its $k!$ orderings where $k = |X|$, say $v_1, v_2, \ldots, v_k$.
For each such $k$-tuple, we consider every possible selection of predecessors $p_1, \ldots, p_k$ and successors $s_1, \ldots, s_k$.
It must hold for every $i$ that $p_i \neq s_i$, $p_i, s_i \in N(v_i)$ and if $s_i \in S$ then $s_i = v_{(i \bmod k) + 1}$ and $p_{(i \bmod k) + 1} = v_i$.
Similarly, if $p_{(i \bmod k)+1} \in S$ then $s_{i} \in S$ for every $i$.

As $G \setminus S$ induces a forest, for every pair $s_i, p_{(i \bmod k)+1}$, there is at most one path between $s_i$ and $p_{(i \bmod k) + 1}$ that does not contain vertices of $S$.
Note that such a selection induces at most one cycle.
Namely, if the paths are disjoint, the selection induces a cycle $C$ which contains $v_i, p_i, s_i$, and the path from $s_i$ to $p_{(i \bmod k) + 1}$ for every~$i$.

Now for every such selection which induces a cycle, we do the following.
For every possible selection of $r - k$ vertices from $V(C) \setminus S$, say $Y$, add the set of connected components induced by $C \setminus (X \cup Y)$ to $\mathcal{P}$.
Note that each such connected component is a path and $\mathcal{P}$ is a family of sets of these paths.
The following shows that the final size of $\mathcal{P}$ and the time required for its construction is polynomial in $n$.



\begin{observation}\label[observation]{obs:higherRedFVS:computePcomplexity}
  Construction of $\mathcal{P}$ can be done in $n^{\Oh(r)}$ time.
\end{observation}
\begin{proof}
  There are no more than $(n + 1)^r$ subsets of $S$ of size $k$, such that $1 \leq k \leq r$.
  For each such subset $X$ of size $k$ we take each of its $k!$ orderings. For each such ordering $v_1,\ldots,v_k$ we fix a predecessor and a successor (taken out of the respective vertex neighborhood) of each $v_i$.
  This constitutes at most $n^{2r}$ different possibilities for each ordering of $X$.
  There is at most one path from the successor of $v_i$ to the predecessor of $v_{(i \mod k) +1}$ in $(G\setminus S)$ as it is a forest.
  As the paths are now fixed, we just need to check whether they form a cycle by checking that the paths are vertex disjoint, which can be done in polynomial time.
  If they do form a cycle $C$, we consider every selection of $r - k$ vertices from $C$, say $Y$ is one of them.
  There is at most $n^r$ such selections of $Y$.
  Now we add to $\mathcal{P}$ the connected components of $C \setminus (X \cup Y)$.
  Note that each such connected component is a path.
  There will be at most $r$ such paths, each can be identified in linear time.
  Altogether, we have spent at most $(n+1)^r r! n^{3r+\Oh(1)} = n^{\Oh(r)}$ time.
\end{proof}

Once the family $\mathcal{P}$ is computed, we solve the above linear program in polynomial time using the Ellipsoid method~\cite{GrotschelLS81} (see also~\cite[Chapter~3]{GLS1988}).
The linear program expresses that for each $X$,$C$, and $Y$ as above, there is ``at least one vertex" selected among $V(C) \setminus (X \cup Y)$.
\begin{equation}\tag{LP$_{s\operatorname{-tuples}}$}\label{lp:higherRedFVS:1}
  \begin{aligned}
    & \text{Minimize} & \sum_{v \in V \setminus S} w(v) \cdot x_v & & \\
    & \text{subject to} & \sum_{v \in \bigcup_{i = 1}^s V(P_i)} x_v &\ge 1 &&&\forall \{P_1, \ldots, P_s\}\in\mathcal{P} \\
    & &  0\le x_v &\le 1  &&&\forall v\in V \setminus S.
  \end{aligned}
\end{equation}

\begin{lemma}\label[lemma]{lem:higherRedFVS:boundOnOPT}
  Let $S^*$ be an $r$-fault tolerant \fvs in $G$ and let $\OPT_{x^*}$ be the optimum value of \eqref{lp:higherRedFVS:1}.
  Then, $\OPT_{x^*} \le w(S^*)$.
\end{lemma}
\begin{proof}
  We claim that the vector $\bm{\widehat{x}}$ defined for $v \in V \setminus S$ as
  \[
    \widehat{x}_v=
    \begin{cases}
      1 &\text{if } v \in S^* \\
      0 & \text{otherwise}
    \end{cases}
  \]
  constitutes a solution to \eqref{lp:higherRedFVS:1}.
  To see this, let $P = \{ P_1, \ldots, P_s \} \in \mathcal{P}$ be an $s$-tuple of paths, let $S$ be the \fvs used in the construction of $\mathcal{P}$ and let $C$ be the cycle formed by $P$ together with $X = \{v_1, v_2, \ldots, v_k\} = S \cap V(C)$ and $Y = \{y_1, y_2, \ldots, y_{r - k}\} = V(C) \setminus S \setminus \bigcup_{i = 1}^s V(P_i)$.
  Note that $|X \cup Y| = r$.

  Since $S^*$ is an $r$-fault tolerant \fvs, we have $|S^* \cap V(C)| \ge r + 1$ and thus $|S^* \cap (V(C) \setminus (X \cup Y))| = \sum_{v \in \bigcup_{i = 1}^s V(P_i)} x_v \ge 1$ as needed.
  Hence, $\OPT_{\bm{x}^*} \le w(S^*)$. 
\end{proof}

Next, based on an optimal solution $\bm{x}^*$ of $\eqref{lp:higherRedFVS:1}$, we create a set of paths $\mathcal{P}_{\bm{x}^*}$.
For each path $P \in \bigcup_{\{P_1,\dots,P_s\} \in \mathcal{P}}\{P_1,\dots,P_s\}$, we include $P $ in $\mathcal{P}_{\bm{x}^*}$ if $\sum_{v \in V(P )} x^*_v \ge \frac{1}{r}$.
Through this process, we are selecting the paths from which we will include at least one vertex in our solution.
Finally, we create an instance $(G \setminus S, w|_{(V\setminus S)},\mathcal{P}_{\bm{x}^*})$ of \multicutInForests.

Consider the corresponding LP relaxation.

\begin{equation}\tag{LP$_{\operatorname{paths}}$}\label{lp:twoRedFVS:2}
\begin{aligned}
& \text{Minimize} & \sum_{v \in V \setminus S} w(v)\cdot y_v & & \\
& \text{subject to} & \sum_{v \in V(P)} y_v &\ge 1 &&&\forall P\in\mathcal{P}_{\bm{x}^*} \\
& &  0\le y_v &\le 1  &&&\forall v\in V \setminus S.
\end{aligned}
\end{equation}

\begin{lemma}\label[lemma]{lem:solutionXBoundOnY}
  Let $\bm{x}^*$ be an optimal solution of \eqref{lp:higherRedFVS:1} and let $\OPT_{\bm{x}^*}$ be its objective value.
  Then, $\bm{y} = \min\{\bm{1},r\bm{x}^*\}$ is a solution to \eqref{lp:twoRedFVS:2}.
  In particular, $\OPT_{\bm{y}^*} \le r \cdot \OPT_{\bm{x}^*}$ holds, where $\OPT_{\bm{y}^*}$ is the value of an optimal solution to \eqref{lp:twoRedFVS:2}.
\end{lemma}
\begin{proof}
  Recall that we have $\sum_{v \in V(P)} x^*_v \ge \frac{1}{r}$ for every path $P \in \mathcal{P}_{\bm{x}^*}$, by the definition of~$\mathcal{P}_{\bm{x}^*}$.
  Thus, we have $\sum_{v \in V(P)} y_v = \sum_{v \in V(P)} \min\{ 1, r \cdot x^*_v\} \ge \min\{ 1, \sum_{v \in V(P)} r \cdot x^*_v\} \ge 1$ for all $P\in\mathcal{P}_{\bm{x}^*}$ and clearly $0 \le y_v \le 1$ for all $v \in V \setminus S$.
  We conclude that $\bm{y}$ is a solution to \eqref{lp:twoRedFVS:2}.
\end{proof}

We now show how to combine the $\alpha$-approximate \fvs with an approximate solution for \multicutInForests to obtain an $(\alpha+r\mu)$-approximate $r$-fault tolerant \fvs.

\begin{lemma}\label[lemma]{lem:twored_approx}
  Let $S$ be an $\alpha$-approximate \fvs and let $\bm{y}$ be an integral solution to \eqref{lp:twoRedFVS:2} of weight at most $\mu$ times the weight of an optimal solution.
  Then, $S'=S \cup \{ v \in V \setminus S \mid y_v = 1 \}$ is an $(\alpha + r\mu)$-approximate $r$-fault tolerant \fvs.
\end{lemma}
\begin{proof}
  Let~$S^*$ be an optimal $r$-fault tolerant \fvs.
  We know that $w(S) \le \alpha \cdot w(S^*)$.
  By \Cref{lem:higherRedFVS:boundOnOPT} there is a solution~$\bm{x}$ to \eqref{lp:higherRedFVS:1} with $\sum_{v \in V \setminus S} w(v)\cdot x_v \le w(S^*)$.
  Thus, by \Cref{lem:solutionXBoundOnY} we have~$\sum_{v \in V \setminus S} w(v)\cdot y_v \le r \mu\cdot w(S^*)$.
  In total we get
  \begin{align*}
    w(S')&=w\left( S \cup \{ v \in V \setminus S \mid y_v = 1 \} \right)\\
    &=
    w(S) + \sum_{v \in V \setminus S} w(v)\cdot y_v \\
    &\le
    \alpha \cdot w(S^*) + r \mu \cdot w(S^*)\\
    &=
    (\alpha+r\mu) w(S^*)
    \,.
  \end{align*}

  We will show that $S'$ is an $r$-fault tolerant \fvs.
  Indeed, if $S$ contains at least $r+1$ vertices in a cycle of the input graph, so does $S'$.
  Thus, we can focus on a cycle~$C$ with $X = V(C) \cap S = \{ v_1, v_2, \ldots, v_k \}$ such that $k \leq r$.

  Suppose for contradiction that $C$ contains at most $r$ vertices of $S'$.
  Let $Y = (S' \cap V(C)) \setminus X$.
  Add arbitrary vertices of $V(C) \setminus (X \cup Y)$ to $Y$ so that $|X \cup Y| = r$.
  During the construction of $\mathcal{P}$, we found $C$ using a certain ordering of vertices in $X$ and with a certain selection of successors and predecessors to those vertices.
  Furthermore we considered $Y$ as the selection of vertices in $V(C) \setminus X$.
  Therefore, we added the set of connected components $P = \{P_1, P_2, \ldots, P_s\}$ of $C \setminus (X \cup Y)$ as a set of paths to $\mathcal{P}$.

  Now consider an optimal solution $\bm{x}^*$ of \eqref{lp:higherRedFVS:1}.
  For at least one path in $P$, say $P_i$, it must hold that $\sum_{v \in V(P_i)} \bm{x}^*_v \geq \frac{1}{r}$ as otherwise $\sum_{v \in \bigcup_{j=1}^s V(P_j)} \bm{x}^*_v~<~1$.
  Therefore in the next step, we include $P_i$ in the constructed instance of \multicutInForests.
  We add the vertices in the solution of this instance to $S'$.
  This ensures that we add to $S'$ at least one vertex from every path selected by \eqref{lp:higherRedFVS:1}.
  From this follows that we added some $v \in P_i$ to $S'$ even though $v \notin X \cup Y$, which is a contradiction.
  Therefore, we have $\left| V(C) \cap S' \right| \ge r + 1$.

  \end{proof}
  \begin{corollary}
    There is a $(2 + r)$-approximation algorithm for unweighted \textsc{$r$-Fault Tolerant \FVS} and $(2 + 2r)$-approximation algorithm for weighted \textsc{$r$-Fault Tolerant \FVS}.
  \end{corollary}
\begin{proof}
We begin with the $2$-approximation algorithm for \textsc{\FVS} by Bafna et al.~\cite{BafnaBF99}.
In polynomial time, we construct \eqref{lp:higherRedFVS:1} and obtain an optimal solution $\bm{x}^*$ for it.
Based on that, we construct $\mathcal{P}_{\bm{x}^*}$ and \eqref{lp:twoRedFVS:2} in polynomial time.
By \Cref{lem:weighted_multicut} or \Cref{lem:optimal_integral} one can in polynomial time find an integral solution to \eqref{lp:twoRedFVS:2} of weight at most $\mu$ times the weight of an optimal solution.
By \Cref{lem:twored_approx} this solution combined with the initial fvs gives $(2+ r \mu)$-approximate $r$-fault tolerant fvs.
The algorithm works in polynomial time as it uses polynomial-time routines.
\end{proof}

%% file: r-ftfvs-hardness.tex
\subsection{Hardness of $r$-Fault Tolerant Feedback Vertex Set}

\DeclarePairedDelimiter\ceil{\lceil}{\rceil}
\DeclarePairedDelimiter\floor{\lfloor}{\rfloor}

\newcommand{\newn}{r + 1 + m(2 \ell + r - 1)}
\newcommand{\mprime}{r + n + m(r + 2\floor{r/2} + 2)}
\newcommand{\mxratio}{\sqrt{(n - 2) / 2}}

\begin{figure}[t]\label[figure]{fig:hardness}
	\centering
	\includegraphics[width=200pt]{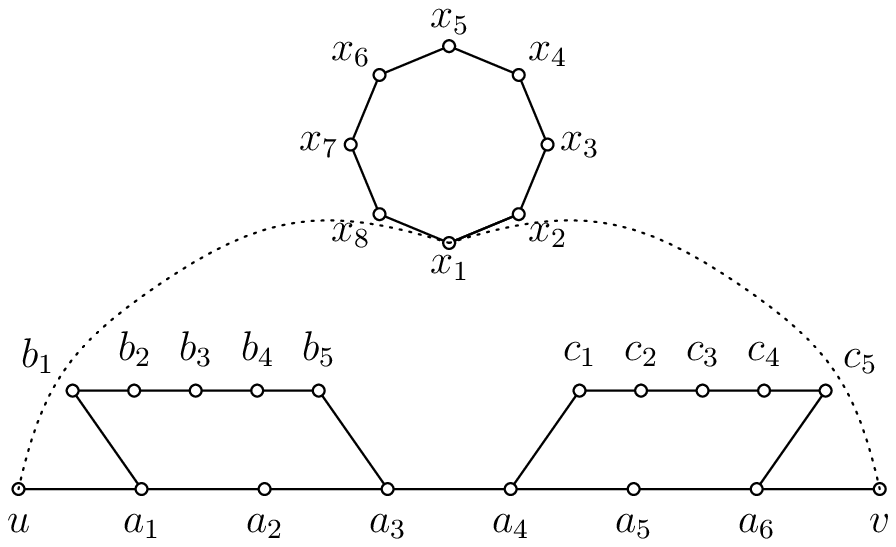}
	\includegraphics[width=200pt]{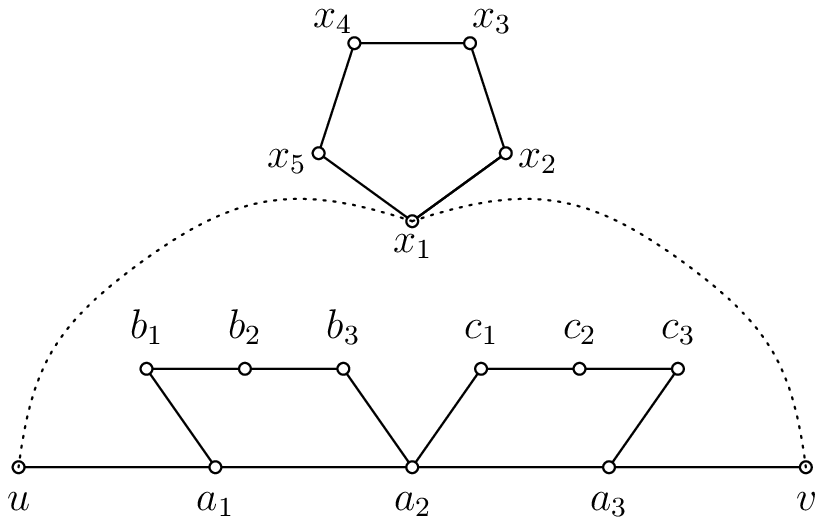}
	\caption{Construction used in \Cref{lem:r-ftfvs-hardness} for $r=7$ and $r=4$.}
\end{figure}

\begin{lemma}\label[lemma]{lem:r-ftfvs-hardness}
  For any fixed $r \geq 0$, it is NP-hard to decide whether for a given graph $G$ and an integer $k$, $G$ contains an $r$-fault tolerant feedback vertex set of size at most $k$.
\end{lemma}
\begin{proof}
  The case of $r = 0$ is equivalent to finding a feedback vertex set of size at most $k$, which is a well known NP-complete problem~\cite{karp1972reducibility}.
The case of $r = 1$ was shown by Misra~\cite{pm-1fvs}.
We extend his approach for $r \geq 2$.

We will show a reduction from \textsc{Vertex Cover}, which is a well known NP-complete problem~\cite{karp1972reducibility}.
  Let~$(G, k)$ be an instance of \textsc{Vertex Cover} and let $G'$ be a graph constructed from $G$ in the following way.
First replace every edge $e = \{u, v\}$ of $G$ by a path $P_e$ with vertices $u, a_1, a_2, \dots, a_{r-1}, v$.
  Additionally create two paths on $\ell = \ceil{r/2}+1$ vertices: $P^b_e$ on $b_1, b_2, \dots, b_\ell$ and $P^c_e$ on $c_1, c_2, \dots, c_\ell$.
  We add edges $\{a_1, b_1\}$, $\{a_{\floor{r/2}}, b_\ell\}$, $\{a_{\ceil{r/2}}, c_1\}$, $\{a_{r-1}, c_\ell\}$.
  Finally add a cycle $X$ with new vertices $x_1, x_2, \dots, x_{r+1}$ and connect $x_1$ to every vertex originally in $V(G)$.
	Two examples for a single edge can be seen in~\Cref{fig:hardness}.

  If $G$ has $n$ vertices and $m$ edges, then this process creates $\newn$ new vertices and $r + 1 + n + m(2l + r + 2)$ new edges.
  As $r$ is fixed, the total size of the construction is bounded by $\mathcal{O}(n + m)$, therefore is polynomial in the size of the input and clearly can be constructed in linear time.

  Let $k' = k + \newn$.
  We will now show that every cycle in $G'$ has size at least $r+1$ and therefore the instance $(G', k')$ of \textsc{$r$-Fault Tolerant Feedback Vertex Set} is not a trivial NO-instance.

  Let $W = V(G') \setminus V(G)$ be the set of vertices added to $G$.
  We will first consider cycles with vertices only in $W$.
  Let $C$ be such a cycle.
  Either $C$ is $X$ or there exists $e \in E(G)$ such that $C$ contains only vertices of $P_e \cup P^b_e \cup P^c_e$.
  The second case leads to only two possible cycles.
  The first cycle consists of vertices $a_1, a_2, \cdots, a_{\floor{r/2}}, b_\ell, b_{\ell - 1}, b_1$, the other one consists of vertices $a_{\ceil{r/2}}, a_{\ceil{r/2} + 1}, \cdots, a_{r-1}, c_\ell, c_{\ell - 1}, c_1$.
  The first cycle has size $\floor{r/2} + \ell = \floor{r/2} + \ceil{r/2} + 1 = r + 1$, the other one $r - \ceil{r/2} + \ell = \floor{r/2} + \ell = r + 1$.

  Let us now consider all cycles containing some vertex from $V(G)$ and excluding $x_1$.
  Let $s~=~|C~\cap~V(G)|$, then $s \geq 3$ as there must exist a cycle $C'$ in $G$ such that $V(C') \subseteq V(C)$. 
  Also note that in $G'$, the distance between any two vertices $u, v$ such that $\{u, v\} \in E(G)$ is exactly $r$.
  Therefore the length of $C$ is at least $s \cdot r \geq r+1$.

  Now any other cycle $C$ containing $x_1$ and some vertex in $V(G)$ must contain at least two vertices from $V(G)$, therefore the size of $C$ is at least $r + 2$.
  Therefore every cycle in $G'$ has size at least $r+1$.

We now show that $G$ has a vertex cover of size at most $k$ if and only if $G'$ has an $r$-fault tolerant feedback vertex set of size at most $k'$.
Suppose that $S'$ is a vertex cover of $G$ of size~$k$.
Then we show that $S = S' \cup W$ is an $r$-fault tolerant feedback vertex set on $G'$.
  Note that $|S| = k + \newn = k'$.

Every cycle with vertices only in $W$ has size exactly $r+1$ and all its vertices are in $S$.
  Let us then consider all cycles which exclude vertex $x_1$ and contain a vertex in $V(G)$.
  Let $C$ be such a cycle.
  As shown above, such $C$ must contain at least $s \geq 3$ vertices from $V(G)$ and therefore has at least $s \cdot r - s \geq r + 1$ in $W$.

  Now any other cycle contains a vertex in $V(G)$ and $x_1$.
  Let $C$ be such a cycle.
  If $C$ contains at least three vertices from $V(G)$, then there are at least $2 \cdot (r-1) + 1 = 2r - 1$ vertices in $W$ and $|V(C) \cap S| \geq 2r - 1 \geq r + 1$.
  Suppose then that $C$ contains exactly two vertices from $V(G)$, say $u$ and $v$.
  Then there are at least $r$ vertices of $C$ in $W$.
  As there is $\{u, v\} \in E(G)$ and $S'$ is a vertex cover of $G$, at least one of $u$ and $v$ must be in $S'$ and therefore $|V(C) \cap S| \geq r + 1$.

  Now let us show that if $G'$ has an $r$-fault tolerant feedback vertex set of size at most $k + \newn$, then there is a vertex cover of size at most $k$ on $G$.
  Let $S$ be an $r$-fault tolerant feedback vertex set of size at most $k + \newn$.
  Consider any cycle on vertices only in $W$.
  Each such cycle has size exactly $r + 1$ and therefore all of its vertices must be in $S$.
  Also every vertex in $W$ is on a cycle with vertices only in $W$, therefore $W \subseteq S$.
  We will show that $S' = S \setminus W$ is a vertex cover of size at most $k$.
  It holds $|W| = \newn$, therefore $|S'| \leq k$.

  Suppose that $S'$ is not a vertex cover.
  Then there is some edge $e = \{u, v\}$ such that $u, v \notin S'$.
  But then consider the cycle $C$ of size $r+2$ consisting of $P_e$ and $x_1$.
  It follows that $|C \cap S| = r$, which contradicts the assumption that $S$ is $r$-fault tolerant.
\end{proof}